\documentclass[aps,letterpaper,twocolumn,prl,showpacs]{revtex4-1}
\usepackage{graphicx,color,textcomp}
\usepackage{mathrsfs,amsmath}   
\usepackage{psfrag,overpic,pst-all}
\usepackage{amssymb}

\begin{document}

\title{Walk-off-induced modulation instability, temporal pattern formation, and frequency comb generation in cavity-enhanced second-harmonic generation}

\author{F. Leo$^1$}
\email{f.leo@auckland.ac.nz}
\author{T. Hansson$^{2,3}$}
\author{I. Ricciardi$^4$}
\author{M. De~Rosa$^4$}
\author{S. Coen$^1$}
\author{S. Wabnitz$^{3,4}$}
\author{M. Erkintalo$^1$}
\email{m.erkintalo@auckland.ac.nz}

\affiliation{$^1$The Dodd-Walls Centre for Photonic and Quantum Technologies, Department of Physics, The University of Auckland, Auckland 1142, New Zealand}
\affiliation{$^2$Department of Applied Physics, Chalmers University of Technology, SE-41296 G\"oteborg, Sweden}
\affiliation{$^3$Dipartimento di Ingegneria dell'Informazione, Universit\`a di Brescia, via Branze 38, 25123 Brescia, Italy}
\affiliation{$^4$CNR-INO, Istituto Nazionale di Ottica, Via Campi Flegrei 34, 80078 Pozzuoli (NA), Italy}

\begin{abstract}
{We derive a time-domain mean-field equation to model the full temporal and spectral dynamics of light in singly resonant cavity-enhanced second-harmonic generation systems. We show that the temporal walk-off between the fundamental and the second-harmonic fields plays a decisive role under realistic conditions, giving rise to rich, previously unidentified nonlinear behaviour. Through linear stability analysis and numerical simulations, we discover a new kind of quadratic modulation instability which leads to the formation of optical frequency combs and associated time-domain dissipative structures. Our numerical simulations show excellent agreement with recent experimental observations of frequency combs in quadratic nonlinear media [Phys. Rev. A \textbf{91}, 063839 (2015)]. Thus, in addition to unveiling a new, experimentally accessible regime of nonlinear dynamics, our work enables predictive modeling of frequency comb generation in cavity-enhanced second-harmonic generation systems. \textcolor{black}{We expect our findings to have wide impact on the study of temporal and spectral dynamics in a diverse range of dispersive, quadratically nonlinear resonators.}
}
\end{abstract}

\maketitle

The generation of frequency combs in high-Q microresonators has attracted significant attention over the last decade~\cite{delhaye_optical_2007}. These combs originate from the third-order optical (Kerr) nonlinearity~\cite{chembo_spectrum_2010}:  four-wave mixing drives modulation instability (MI), leading to the growth of signal and idler sidebands~\cite{hansson_dynamics_2013}. In the time domain, such ``Kerr'' combs can correspond to temporal dissipative patterns or localised structures --- temporal cavity solitons (CSs)~\cite{wabnitz_suppression_1993, leo_temporal_2010,coen_modelling_2013}.

Certain microresonators exhibit a weak second-order $\chi^{(2)}$ nonlinearity, which may lead to the intracavity conversion of the Kerr comb to shorter wavelengths~\cite{miller_on-chip_2014, jung_green_2014}. But recent experiments in free-space resonators have remarkably demonstrated that frequency combs can also arise entirely through $\chi^{(2)}$ effects. On the one hand, it is known that cascaded second-harmonic generation (SHG), subject to large phase mismatch, can give rise to an effective Kerr nonlinearity~\cite{desalvo_self_1992, stegeman_chi2_1996}, and comb generation has indeed been observed under such conditions~\cite{ulvila_frequency_2013, ulvila_high-power_2014}. But on the other hand, frequency combs have recently been observed also for \emph{phase-matched} cavity-enhanced SHG~\cite{ricciardi_frequency_2015}.

In cavity-enhanced SHG, comb formation initiates from the spontaneous down-conversion of the second-harmonic field~\cite{ricciardi_frequency_2015}, a process widely investigated in the context of internally pumped optical parametric oscillators (OPOs)~\cite{schiller_subharmonic_1996,schiller_subharmonic_1993,   lodahl_pattern_1999, lodahl_spatiotemporal_2001}. However, theoretical analyses have hitherto been limited to a very small number of frequency components: no models have been put forward that would allow for the full frequency comb dynamics to be examined. For quadratically nonlinear, \emph{spatially diffractive} cavities full models do exist, and their study has revealed numerous \emph{spatial} phenomena whose \emph{temporal} analogs would be associated with frequency combs, including pattern formation~\cite{etrich_pattern_1997,lodahl_pattern_1999, lodahl_pattern_2000} and quadratic cavity solitons~\cite{etrich_solitary_1997}. Unfortunately, insights obtained in the spatial domain are of limited use in the temporal domain. This is because the very large temporal walk-off characteristic to quadratically interacting fields is unique to the time-domain; the corresponding spatial walk-off is weak~\cite{ward_transverse_1998,skryabin_walking_2001} and typically neglected altogether~\cite{etrich_pattern_1997,lodahl_pattern_1999, lodahl_pattern_2000, etrich_solitary_1997, fuerst_spatial_1997}. Although MI-induced pulse train generation has been explored in dispersive quadratic resonators~\cite{trillo_pulse_1996}, these investigations also overlook the role of temporal walk-off. Studies of quadratically nonlinear single-pass (non-cavity) systems have included the full effects of walk-off~\cite{buryak_optical_2002, bache_nonlocal_2007,conforti_nonlinear_2010}, however, it is well-known that dissipative systems (including driven cavities) differ fundamentally from conservative (single-pass) systems~\cite{haelterman_dissipative_1992, akhmediev_dissipative_2008}. Thus, given the recent observations in \emph{cavity} SHG systems~\cite{ulvila_frequency_2013, ulvila_high-power_2014, ricciardi_frequency_2015}, as well as progresses in the fabrication of high-Q microresonators with strong $\chi^{(2)}$ response~\cite{furst_naturally_2010, lin_wide-range_2013, kuo_second_2014}, there is clearly a need to advance the theoretical understanding of temporal dynamics in dispersive quadratic resonators.

\looseness=-1 In this Letter, we introduce a framework that enables quantitative modelling of the temporal and spectral dynamics in dispersive, quadratically nonlinear resonators. Our general approach can be applied to many configurations of interest, but here we focus on singly-resonant cavity-enhanced SHG where frequency combs have been reported~\cite{ricciardi_frequency_2015}. We derive a single time-domain mean-field equation that allows for the system's full spectral and temporal dynamics to be described. We find that, not only is it inappropriate to neglect temporal walk-off, but that it fundamentally underpins the dynamics. In particular, our study reveals a new kind of walk-off-induced cavity MI, that triggers the formation of optical frequency combs and corresponding dissipative temporal patterns. Our simulation results agree with experimental observations~\cite{ricciardi_frequency_2015}.

\looseness=-1 We consider the evolution of the slowly-varying envelopes $A(z,\tau)$ and $B(z,\tau)$ of electric fields centered at frequencies $\omega_0$ and $2\omega_0$, respectively, in a resonator containing a dispersive quadratic medium~\cite{ferro_periodical_1995}. The resonator is driven with a continuous wave (cw) field $A_\mathrm{in}$ at $\omega_0$, injected into the cavity through a coupler with power transmission coefficients $\theta_1$ and $\theta_2$ at frequencies $\omega_0$ and $2\omega_0$, respectively. The intracavity fields $A_{m+1}(0,\tau)$ and $B_{m+1}(0,\tau)$ at the beginning of the $(m+1)^{\mathrm{th}}$ roundtrip can be related to the fields $A_{m}(L,\tau)$ and $B_{m}(L,\tau)$ at the end of the $m^{\mathrm{th}}$ roundtrip as:
\begin{align}
A_{m+1}(0,\tau) &= \sqrt{1-\theta_1}A_{m}(L,\tau)e^{-i\delta_1}+\sqrt{\theta_1}A_\mathrm{in}\label{boundary_fun}\\
B_{m+1}(0,\tau) &= \sqrt{1-\theta_2}B_{m}(L,\tau)e^{-i\delta_2}.\label{boundary_SH}
\end{align}
Here $L$ is the length of the medium and $\delta_{1,2}$ represent the phase detunings of the intracavity fields from the cavity resonances closest to $\omega_0$ and $2\omega_0$, respectively.

Assuming that diffraction \textcolor{black}{and higher-order nonlinearities} can be neglected, the evolution of the fields through the \textcolor{black}{quadratically} nonlinear medium obey the coupled equations~\cite{buryak_optical_2002}\textcolor{black}{\footnote{\textcolor{black}{When modelling ultra-broadband signals, it may be necessary to use more complex generalized envelope equations~\cite{conforti_nonlinear_2010}.}}}
\begin{align}
&\hspace{-3pt}\frac{\partial A_m}{\partial z} =\left[-\frac{\alpha_{c1}}{2}- i\frac{{k}_1''}{2}\frac{\partial^2}{\partial \tau^2}\right]\hspace{-2pt} A_m+i\kappa B_mA_m^*e^{-i \Delta k z}, \label{fundamental_K}\\
&\hspace{-3pt}\frac{\partial B_m}{\partial z} =\left[-\frac{\alpha_{c2}}{2} - \Delta {k}'\frac{\partial }{\partial \tau}-i\frac{{k}_2''}{2}\frac{\partial^2 }{\partial \tau^2}\right]\hspace{-2pt} B_m+i\kappa A_m^2 e^{i\Delta k z},\label{SH_K}
\end{align}
where $\alpha_{c1,2}$ describe propagation losses, $\Delta k = 2k(\omega_0)-k(2\omega_0)$ is the wave-vector mismatch associated with the SHG process $\omega_0+\omega_0 = 2\omega_0$, $\Delta {k}' = \mathrm{d}k/\mathrm{d}\omega|_{2\omega_0}-\mathrm{d}k/\mathrm{d}\omega|_{\omega_0}$ is the corresponding group-velocity mismatch, and ${k}''_{1,2} = \mathrm{d}^2k/\mathrm{d}\omega^2|_{\omega_0, 2\omega_0}$ are the group-velocity dispersion coefficients. \textcolor{black}{(Including higher-order dispersion is straightforward but omitted for brevity.)} The nonlinear coupling strength is described by $\kappa$, normalised such that $|A_m|^2$, $|B_m|^2$  and $|A_\mathrm{in}|^2$ are measured in Watts. Specifically, $\kappa = \sqrt{8}\omega_0\chi^{(2)}_\mathrm{eff}/\sqrt{c^3n_1^2n_2\epsilon_0}$, where $\chi^{(2)}_\mathrm{eff}$ is an effective second-order susceptibility (measured in $\mathrm{V}^{-1}$) that includes the spatial overlap between the interacting fields~\cite{buryak_optical_2002}, $c$ is the speed of light, $n_{1,2}$ are the refractive indices at $\omega_0$ and $2\omega_0$, respectively, and $\epsilon_0$ is the vacuum permittivity.

\looseness=-1 Equations~\eqref{boundary_fun}--\eqref{SH_K} allow the temporal and spectral dynamics in cavity-enhanced SHG systems to be modelled. Depending on $\theta_{1,2}$, the physical system can be singly- or doubly-resonant.  Changing the driving term to Eq.~\eqref{boundary_SH} allows for the modelling of degenerate OPOs~\cite{longhi_ultrashort_1995}. Adding a third equation would expand the framework to non-degenerate OPOs. All these different configurations are associated with unique dynamics and they allow for different ways to simplify the general model equations; considering all possibilities is beyond the scope of this Letter.

Here we focus on an SHG cavity that is resonant only for wavelengths around the pump~\cite{ricciardi_frequency_2015}, and we thus set $\theta_2 = 1$~\cite{lodahl_pattern_1999}. Moreover, we assume that the cavity has high finesse around $\omega_0$, such that $A_m(z,\tau)$ remains approximately constant over one roundtrip. By treating the dispersion at the second-harmonic as a small perturbation, it is possible to combine Eqs.~\eqref{boundary_fun}--\eqref{SH_K} into a single mean-field equation~\cite{lodahl_pattern_1999, lodahl_pattern_2000}. This is not a valid approximation here, however, due to the large walk-off term $\Delta{k}'$. Notwithstanding, we have derived a more general mean-field equation that is valid also for large $\Delta{k}'$. Briefly, Eq.~\eqref{SH_K} is first solved in the Fourier domain~\cite{bache_nonlocal_2007}:
\begin{equation}
\mathscr{F}\left[B_m(z,\tau)\right] \approx \kappa\mathscr{F}\left[A_m^2(z,\tau)\right]\displaystyle\frac{e^{i\Delta k z}-e^{\hat{k}z}}{\Delta k+i\hat{k}},
\label{slaved_SH}
\end{equation}
where we used $B_m(0,\tau) = 0$, $\mathscr{F}\left[\cdot\right]=\int_{-\infty}^{\infty}\cdot\,e^{i\Omega\tau}\,\mathrm{d}\tau$ denotes Fourier transformation, and $\hat{k}(\Omega) = -\alpha_{c,2}/2+i\left[\Delta{k}'\Omega+({k}_2''/2)\Omega^2\right]$. We then substitute $B_m(z,\tau)$ into Eq.~\eqref{fundamental_K}, integrate over one roundtrip keeping $A_m(z,\tau)$ constant, and combine the result with Eqs.~\eqref{boundary_fun} and ~\eqref{boundary_SH}. This yields a single mean-field equation:
\begin{align}
  \label{MF}
  t_\mathrm{R}\frac{\partial A(t,\tau)}{\partial t} =& \bigg[ -\alpha_1 - i \delta_1 -iL\frac{{k}_1''}{2}\frac{\partial^2}{\partial \tau^2} \bigg]A \nonumber \\
  & - \rho A^*\left[A^2(t,\tau)\otimes I(\tau)\right] + \sqrt{\theta_1}\,A_\mathrm{in}.
\end{align}
Here, $t$ is a ``slow time'' variable, linked to the roundtrip index as $A(t=mt_\mathrm{R},\tau) = A_m(z=0,\tau)$~\cite{leo_temporal_2010, coen_modelling_2013, haelterman_dissipative_1992}, $t_\mathrm{R}$ is the cavity roundtrip time, $\alpha_1 = (\alpha_{c1}L+\theta_1)/2$, $\rho = (\kappa L)^2$, $\otimes$ denotes convolution and the nonlinear response function $I(\tau) = \mathscr{F}^{-1}[\hat{I}(\Omega)]$, with $\hat{I}(\Omega) = \left[(1-e^{-ix}-ix)/x^2\right]$ and $x(\Omega) = \left[\Delta k+i\hat{k}(\Omega)\right]L$.

\begin{figure}[t]
  \centering
  \includegraphics[width = 8.3cm]{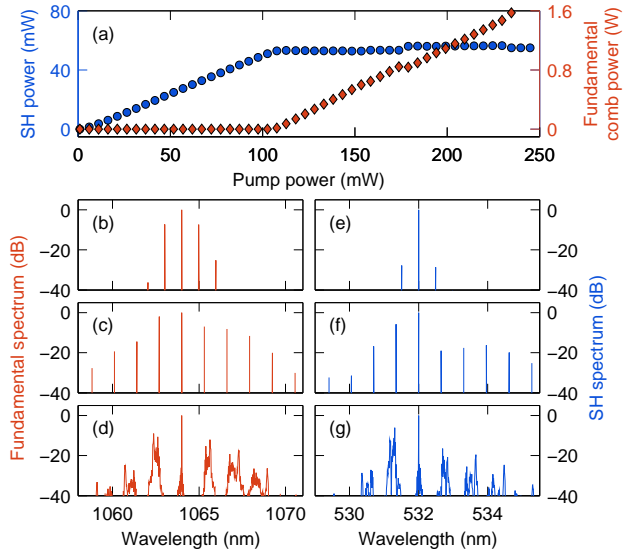}
  \caption{Numerically simulated frequency comb generation in singly-resonant, phase-matched cavity SHG. (a) Evolution of power contained in the second-harmonic (SH, blue circles) and in the frequency comb around the fundamental frequency (red diamonds) as a function of pump power for $\delta_1 = 0$. (b--d) Examples of simulated frequency combs around the fundamental wavelength for different detunings and pump powers. In (b) $\delta_1 = 0.004$, $P_\mathrm{in} = 170$~mW; in (c) $\delta_1 = -0.02$, $P_\mathrm{in} = 2$~W; and in (d) $\delta_1 = -0.01$, $P_\mathrm{in} = 2$~W. (e--g) Corresponding spectra around the second-harmonic wavelength.}
  \label{Fig1}
\end{figure}

\looseness=-1 The mean-field Eq.~\eqref{MF} reveals an effective third-order nonlinearity, in line with the truncated three-wave analysis in~\cite{ricciardi_frequency_2015}. In general, the nonlinear response $I(\tau)$ is non-instantaneous, akin to those reported in studies of non-cavity, \emph{single-pass} SHG ~\cite{bache_nonlocal_2007, nikolov_quadratic_2003}. Equation~\eqref{MF} however introduces such responses into the description of \emph{cavity} SHG, enabling the analysis of new classes of phenomena. As expected, $\mathrm{Im}[\hat{I}(0)] = 0$ for $\Delta k = 0$, indicating \textcolor{black}{that phase-matched SHG does not give rise to} Kerr-like instantaneous phase shifts. Conversely, for $|\Delta k|\gg |\hat{k}(\Omega)|$ almost pure Kerr-nonlinearity ensues, with Eq.~\eqref{MF} reducing to the AC-driven nonlinear Schr\"odinger equation describing, e.g., coherently-driven Kerr cavities~\cite{wabnitz_suppression_1993, leo_temporal_2010, coen_modelling_2013, lugiato_spatiotemporal_1987}. Because Eq.~\eqref{MF} includes effects of dispersion and models the evolution of the full temporal electric field envelope $A(t,\tau)$, it intrinsically takes all relevant frequency components into account (see also ~\cite{coen_modelling_2013}), including those that arise from the spontaneous down-conversion of the second-harmonic field~\cite{ricciardi_frequency_2015, schiller_subharmonic_1996, lodahl_spatiotemporal_2001}. This should be contrasted with prior approaches that have used coupled mode equations to investigate a limited number of frequency components only~\cite{ricciardi_frequency_2015, lodahl_pattern_1999, lodahl_spatiotemporal_2001}. \textcolor{black}{Such a coupled-mode system can be recovered by injecting into Eq.~\eqref{MF} a truncated ansatz that only contains a small number of Fourier modes.}

To show that the above framework allows for the modelling of realistic systems, we numerically integrate Eq.~\eqref{MF} with pump-resonator parameters approximating the phase-matched ($\Delta k = 0$), \textcolor{black}{continuously-driven} SHG configuration in~\cite{ricciardi_frequency_2015}. (The full map given by Eqs.~\eqref{boundary_fun}--\eqref{SH_K} yields almost identical results.) The resonator has a finesse of $\mathcal{F} = \pi/\alpha_1 = 160$ at the pump wavelength of 1064~nm, and it contains a nonlinear crystal with length $L=15~\mathrm{mm}$ (passed through once per roundtrip). For simplicity, we assume $\alpha_{c1} =\alpha_{c2}$ and that the resonator is critically coupled ($\alpha_1=\theta_1$). The nonlinear coupling coefficient $\kappa = 11.14~\mathrm{W^{-1/2}m^{-1}}$ was estimated from the SHG efficiency measured in~\cite{ricciardi_frequency_2015}, and the group-velocity dispersion coefficients ${k}_1'' = 0.234~\mathrm{ps^2/m}$ and ${k}_2'' = 0.714~\mathrm{ps^2/m}$ were evaluated using the Sellmeier equation for lithium niobate~\cite{gayer_temperature_2008}. This also yielded the temporal walk-off coefficient $\Delta{k}' = 792~\mathrm{ps/m}$. The detuning $\delta_1$ is not straightforward to extract from experiments, and we thus treat it as a free parameter. Our simulations begin from noise, and we iterate each simulation for at least 40000 roundtrips to ensure the dismissal of transients (photon lifetime is about 26 roundtrips). To reduce computational burden, our simulations use a smaller (about 200~ps) time window than the 2~ns cavity roundtrip time of the resonator in~\cite{ricciardi_frequency_2015}. Physically, this corresponds to neglecting portions of free-space propagation, which does not influence the results.

\looseness=-1 A distinctive feature observed in~\cite{ricciardi_frequency_2015} was the clamping of the second-harmonic power at the onset of comb formation. This behaviour is reproduced in our simulations, as shown in Fig.~\ref{Fig1}(a). Here we plot the second harmonic power (blue circles) as well as the power contained in all the comb lines around the pump (red diamonds) as a function of pump power $P_\mathrm{in} = |A_\mathrm{in}|^2$. The saturation of the second-harmonic is clearly correlated with the emergence of a frequency comb around $\omega_0$. Selected examples of simulated spectra are shown in Figs.~\ref{Fig1}(b)--(d) for various pump powers and detunings (see caption). Of course, because the second-harmonic field is slaved to the fundamental [see Eq.~\eqref{slaved_SH}], frequency combs also manifest themselves around $2\omega_0$, as shown in Figs.~\ref{Fig1}(e)--(g). We note that the overall agreement between these simulation results and the experimental observations in~\cite{ricciardi_frequency_2015} is remarkably good.

\begin{figure}[b]
  \centering
  \includegraphics[width = 8.3cm,clip=true]{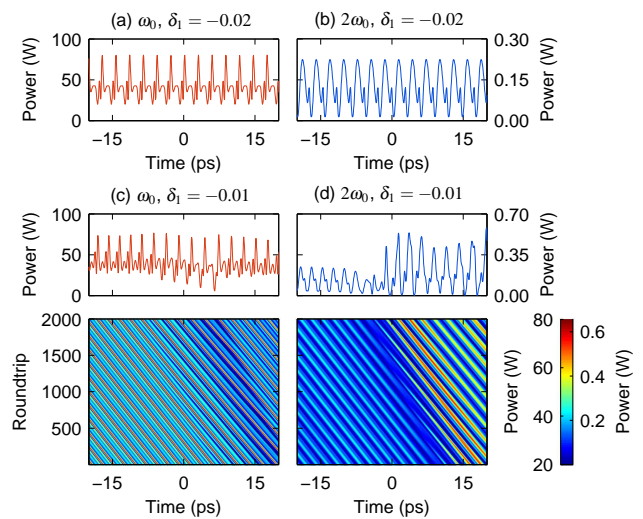}
  \caption{Numerically simulated temporal patterns at the (a, c) fundamental and (b, d) second-harmonic wavelengths for $P_\mathrm{in} = 2~\mathrm{W}$. Density maps show the evolution of the patterns over 2000 consecutive roundtrips, revealing a temporal drift that is much smaller than the walk-off parameter. In (a, b) $\delta_1 = -0.02$, the patterns are fully stable and correspond to spectra in Fig.~\ref{Fig1}(c, f). In (c, d) $\delta_1 = -0.01$ and the patterns exhibit minor evolution over very slow time scales. Corresponding spectra are shown in Fig.~\ref{Fig1}(d) and (g).}
  \label{Fig2}
\end{figure}

The combs in Figs.~\ref{Fig1}(b) and (c) are fully stable and phase-locked: they correspond to periodic temporal patterns in the time domain. Example intensity profiles, corresponding to the fundamental and second-harmonic spectra in Figs.~\ref{Fig1}(c) and (f), are shown in Figs.~\ref{Fig2}(a) and (b), respectively. The patterns are complex, yet stable. They exhibit, however, a constant temporal drift of about $-0.93~\mathrm{ps/m}$ in the reference frame of Eq.~\eqref{MF}, which moves with the group velocity of light at $\omega_0$. Despite their noisy appearance, the combs in Figs.~\ref{Fig1}(d) and (g) also exhibit a high degree of stability. This can be appreciated from Figs.~\ref{Fig2}(c) and (d), where we show the corresponding temporal profiles and their evolution over 2000 roundtrips. The fields only evolve slowly, barring again a trivial temporal drift of about $-0.85~\mathrm{ps/m}$. Although a detailed discussion is beyond the scope of our study, we remark that (i) the temporal drifts are about three orders of magnitude smaller than the walk-off parameter $\Delta {k}'$ and that (ii) despite the large $\Delta {k}'$, the fundamental and the second-harmonic patterns drift at the same rate.

\begin{figure}[t]
  \centering
  \includegraphics[width = 8.3cm]{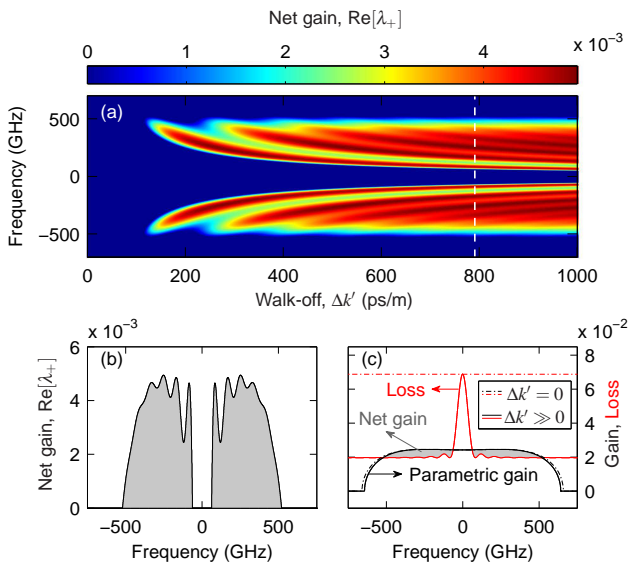}
  \caption{(a) MI gain for a singly-resonant SHG system as a function of frequency and walk-off. Other parameters are as in Fig.~\ref{Fig1}(b). White-dashed line highlights the walk-off value corresponding to simulations in Fig.~\ref{Fig1} and (b) shows the corresponding gain profile. (c) Parametric gain (black curve) and total sideband losses (red curve) for zero walk-off (dash-dotted curves) and for a large walk-off $\Delta{k}' = 792~\mathrm{ps/m}$ (solid curves). Shaded areas show net gain for $\Delta{k}' = 792~\mathrm{ps/m}$.}
  \label{Fig3}
\end{figure}

\looseness=-1 In our simulations, comb formation initiates from an MI-like growth of sidebands around the pump. To gain more insight, we have studied the stability of the system's steady-state cw mode $A_{0}$ against periodic perturbations. The cw mode can be found by setting all the derivatives in Eq.~\eqref{MF} to zero and noting that, for a cw field, the convolution $A_{0}^2\otimes I(\tau) = A_{0}^2 \hat{I}(0)$. This yields $A_{0} = \sqrt{\theta_1}A_\mathrm{in}/[\alpha_1+i\delta_1+\rho \hat{I_0}(0) P_{0}]$, where the cw intracavity power $P_{0} = |A_{0}|^2$ satisfies the cubic polynomial
\begin{equation}
\theta_1P_\mathrm{in} =\rho^2P_{0}^3|\hat{I}(0)|^2+2\rho(\alpha_1R+\delta_1J) P_{0}^2+(\alpha_1^2+\delta_1^2)P_{0},
\end{equation}
with $R = \mathrm{Re}[\hat{I}(0)]$ and $J = \mathrm{Im}[\hat{I}(0)]$. To examine the stability of $A_0$, we inject into Eq.~\eqref{MF} the ansatz ${A = A_0+a_1\exp(\lambda\, t/t_\mathrm{R}+i\Omega\tau)+a_2\exp(\lambda^*\,  t/t_\mathrm{R}-i\Omega\tau)}$. At first order in $a_1,\, a_2$ we obtain a linear system whose eigenvalues $\lambda$ read
\begin{widetext}
\begin{equation}
\lambda_\pm = -\left(\alpha_1+\rho P_0[\hat{I}(\Omega)+\hat{I}^*(-\Omega)]\right) \pm \sqrt{|\hat{I}(0)|^2\rho^2P_0^2-\left(\delta_1-\frac{{k}_1''L}{2}\Omega^2-i\rho P_0[\hat{I}(\Omega)-\hat{I}^*(-\Omega)]\right)^2}.
\label{MIgain}
\end{equation}
\end{widetext}
\looseness=-1 Physically, growth of sidebands at $\omega_0\pm \Omega$ occurs via parametric down-conversion of the second-harmonic at $2\omega_0$~\cite{schiller_subharmonic_1993,  schiller_subharmonic_1996, lodahl_pattern_1999, lodahl_spatiotemporal_2001}; the square root term in Eq.~\eqref{MIgain} describes the corresponding gain. But for net MI growth [$\mathrm{Re}(\lambda_+)>0$], parametric gain must overcome both linear ($\propto \alpha_1$) and nonlinear ($\propto \mathrm{Re}[\hat{I}(\Omega)+\hat{I}^*(-\Omega)]$) losses. Notably, the latter arise from competing sum-frequency processes that can convert the $\omega_0$ sidebands back to the vicinity of $2\omega_0$ ($\hat{I}(\Omega)$ describes the efficiency of creating photons at $2\omega_0+\Omega$).

\looseness=-1 Figure~\ref{Fig3}(a) shows the MI gain spectrum computed for the same parameters as in Fig.~\ref{Fig1}(b, e), but for a wide range of walk-off parameters $\Delta {k}'$. The value corresponding to simulations in Fig.~\ref{Fig1} is indicated with the dashed vertical line, and Fig.~\ref{Fig3}(b) shows the corresponding gain profile in more detail. It is clear that MI indeed does not occur for $\Delta {k}'=0$, in agreement with studies in spatial systems~\cite{lodahl_pattern_1999, lodahl_pattern_2000}. But for sufficiently large walk-offs, MI does occur. This can be understood by noting that a large $\Delta {k}'$ causes sum-frequency processes to be heavily phase-mismatched, thereby reducing the nonlinear losses experienced by sidebands around $\omega_0$. We show this explicitly in Fig.~\ref{Fig3}(c), where we compare the total sideband losses ($\alpha_1+\rho P_0\mathrm{Re}[\hat{I}(\Omega)+\hat{I}^*(-\Omega)]$) for zero (red dash-dotted curve) and large (red solid curve) walk-off. Also shown are (black curves) the corresponding parametric down-conversion gain profiles [real part of the square root term in Eq.~\eqref{MIgain}]. The parametric gain does not materially depend on walk-off; it is the reduction in sum-frequency losses that enables MI. Figure~\ref{Fig3}(c) also illustrates that the several local maxima (minima) of the MI gain spectra [see e.g. Fig.~\ref{Fig3}(b)] arise from the minima (maxima) of the $\mathrm{sinc}^2$-shaped nonlinear loss spectrum. Thus, in addition to unveiling a new class of walk-off-induced quadratic cavity MI and enabling MI gain spectra to be analytically evaluated, Eq.~\eqref{MIgain} formally confirms the conclusions in~\cite{ricciardi_frequency_2015}.

\looseness=-1 To the best of our knowledge, our work represents the first quantitative theoretical study of the full temporal and spectral dynamics in a realistic quadratically nonlinear resonator. Starting from a general theoretical framework, we have derived a single mean-field equation that allows singly-resonant SHG systems to be accurately modelled, yielding simulation results in remarkable agreement with experimental observations~\cite{ricciardi_frequency_2015}. Our analysis reveals that the cavity MI, which triggers the formation of frequency combs and drifting temporal patterns, is fundamentally underpinned by the large temporal walk-off between the interacting fields. We emphasize that our findings are not unique to the particular parameters used. Besides different singly-resonant SHG systems, preliminary analyses based on the general framework outlined in our work show that walk-off induced cavity MI can manifest itself in many different quadratic resonator configurations, giving rise to rich and unexplored dynamics. Thus, in addition to enabling predictive modelling of quadratic frequency combs, our work opens up a new regime of nonlinear dynamics for exploration. Finally, the fact that walk-off induces MI through an effective non-instantaneous third-order nonlinearity [see Eq.~\eqref{MF}] suggests interesting links to other physical systems, such as single-pass nonlinear generation schemes, spatially diffractive cavities, Bose-Einstein condensates and fluid dynamics, where analogous non-local nonlinearities~\cite{bache_nonlocal_2007, nikolov_quadratic_2003, kroliowski_modulational_2001} and walk-off~\cite{bache_nonlocal_2007, onorato_modulational_2006} have also been shown to impact MI, and even give rise to soliton formation.

\begin{acknowledgments}
 We are grateful to Gian-Luca Oppo for useful discussions. We acknowledge support from the Marsden Fund of The Royal Society of New Zealand, the Swedish Research Council (grant no. 2013-7508), the Finnish Cultural Foundation and the Italian Ministry of University and Research (grant no. 2012BFNWZ2 and Progetto Premiale QUANTOM -- Quantum Opto-Mechanics).
\end{acknowledgments}

\end{document}